\documentclass{XrU2005}
\usepackage{graphicx,subfigure,epsf}
\title{FRII sources at $z>0.5$: X-ray properties of the core and extended emission}
\author[1]{E. Belsole}
\author[2,3]{J.H. Croston}
\author[1]{D.M. Worrall}
\author[3]{M. J. Hardcastle}
\affil[1]{H.H. Wills Phyisics Laboratory, University of Bristol, Tyndall Avenue, Bristol BS8 1TL, UK}
\affil[2]{Service d'Astrophysique, bat 709 CEA- Saclay, L'orme des Merisiers, 91191 Gif sur Yvette Cedex, France }
\affil[3]{School of Physics, Astronomy and Mathematics, University of
Hertfordshire, College Lane, Hatfield, Hertfordshire AL10 9AB, UK}

\def\xmm{{\it XMM-Newton\/}}
\def\cha{{\it Chandra\/}}
\def \nh {$N_{\rm H}$}

\begin{document}

\keywords{active galaxies, radio galaxies,  high redshift, quasars,
general, X-ray, synchrotron, inverse Compton}

\maketitle
\begin{abstract}
Active galaxies are the most powerful engines in the Universe for
converting gravitational energy into radiation, and their study at all
epochs of evolution is therefore important.  Powerful radio-loud
quasars and radio galaxies have the added advantage that, since their
radio jets need X-ray-emitting gas as a medium in which to propagate,
the sources can be used as cosmological probes to trace significant
atmospheres at high redshift.  The radio emission can be used as a
measure of source orientation, and sensitive X-ray measurements,
especially when used in combination with multi-wavelength data, can be
used to derive important results on the physical structures on a range
of sizes from the cores to the large-scale components. In this paper
we present
new results on a significant sample of powerful radio galaxies and
quasars at $z > 0.5$, drawn from the 3CRR catalogue and selected to
sample a full range of source orientation. Using high-quality
observations from \xmm\ and \cha, we discuss the X-ray
properties of the cores, jets, lobes and cluster gas, and, through the
incorporation of multi-wavelength data, draw conclusions about the
nature of the emission from the different components.
\end{abstract}

\section{Introduction}
Powerful ($P_{\rm 178 MHz} > 5\times10^{24}$ W s$^{-1}$ Hz$^{-1}$)   radio sources are
visible across the Universe and can thus be used to probe a number of
physical conditions at high redshift, from accretion processes to
Active Galactic Nucleus (AGN) environments at early epochs. On small
scales, X-ray emission from the central AGN can be used to probe  the
process of converting gravitational energy into radiation, as spectral
and variability studies have shown that at least some of the X-ray
emission comes from regions very close to the central engine. At least
part of the X-ray
emission  is expected to be anisotropic as a result
of relativistic beaming and anisotropic absorption. In particular,
Unification Models explain the observed differences among AGNs as the
result of their orientation with respect to
the line of sight (e.g. \citealt{barthel89, UrryPadovani95}), and an optically thick obscuring torus is invoked to hide the nucleus of objects (namely radio galaxies) viewed at large
angles to the jet axis (e.g. \citealt{barthel89}).

To understand the accretion mechanism(s) we need to understand
orientation effects, especially in X-ray selected samples.
In the simple picture of an obscuring torus, quasar light  heats the gas and dust of the torus and 
thermal radiation is re-emitted isotropically in the mid/far-infrared in
order to maintain energy balance in the inner regions. Thus, in
principle, far-infrared radiation should provide an
orientation-independent measure of the emitted power
from the central engine.  The sensitivity in the mid/far-infrared of the {\em Spitzer} 
satellite provides for the first time
the possibility to test this hypothesis for a large number of
objects.  

On larger spatial scales, extended X-ray emission is observed from
spatial regions coincident with the radio lobes, and the combination of
the X-ray and radio radiation can be used to measure the particle
content and magnetic field in the radio lobes. 
In addition, the well collimated
relativistic jets associated with these sources, require a medium in
which to propagate. As a result powerful high redshift radio galaxies are
potential tracers of the  formation and evolution of the most massive
galaxies and clusters (e.g., \citealt{cf03, hw99, dmw01}).
Powerful radio sources
are thus key objects to understand the cosmological evolution of
accretion/radiation mechanisms, relativistic effects and plasma
physics in the early Universe (e.g., \citealt*{brunetti01, hbw01,
marshall05, overzier05}). 

In this paper we present high-quality X-ray observations obtained with
\cha\ and \xmm\ of 19 sources in the redshift range $0.5<z<1.0$, which are mostly part of a larger sample of Faranoff-Riley type II (FRII) radio galaxies and quasars currently being  observed with {\em
Spitzer}. We discuss the properties of the core and particle content of the radio lobes and we  present preliminary results on the cluster-like environment of these sources.

Throughout this paper we use the concordance cosmology with $h_0 =
H_0/100\ {\rm km}\ {\rm s}^{-1}\ {\rm Mpc}^{-1}
= 0.7$, $\Omega_{\rm M} = 0.3$, $\Omega_{\Lambda} = 0.7$. If not
otherwise stated, quoted errors are $1\sigma$ for one interesting
parameter.

\section{The sample}
Low-radio-frequency optically-thin synchrotron radiation from the radio
lobes of radio-loud sources should be isotropic.
Thus, selection of objects via their 
low-frequency radio emission represents the most reliable  method for
selecting an orientation-unbiased sample. 
The sample discussed in this paper is composed of 19 sources at  redshift $0.5<z<1.0$,  selected from the 3CRR catalogue  \citep*{3crrcat} at 178 MHz, and having \cha\ or \xmm\ observations. This work does not aim at statistically testing  unification models, for which a random selection from the parent
sample would be necessary. Instead we intend to look for
differences in the X-ray emitting components
between quasars and radio galaxies, as would be
expected in unification schemes. The sources are also being observed
with {\em Spitzer} and their selection as part of a larger sample
 was based on convenient scheduling of {\em Spitzer} observations.

\cha\ or \xmm\ observations of 9 sources (3C\,184, 3C\,200, 3C\,220.1, 3C\,228, 3C\,263,
3C\,275.1, 3C\,292, 3C\,330, 3C\,334) were awarded to us in
  support of various projects, while data for the remaining sources were extracted from the \cha\ archive.
Table \ref{tab:sources} lists the sources used in this work. 
\begin{table*}
\begin{center}
\caption{The sample. 
Galactic column density is from \citet{dlnh}; NRLG means Narrow Line
Radio Galaxy; LERG means low-excitation radio galaxy. Redshifts and
positions are taken from \cite{3crrcat}}\vspace{1em}
    \renewcommand{\arraystretch}{1.2}
\label{tab:sources}
\begin{tabular}[h]{l|cclclc}
\hline
Source & RA(J2000) & Dec(J2000) & redshift  & scale &type & \nh \\
        & $^{\rm h~m~s}$ &$^{\circ~\prime~\prime\prime}$ & &
        arcsec/kpc & & 10$^{20}$ cm $^{-2}$\\
\hline
3C\,6.1 &  00 16 30.99 & +79 16 50.88  & 0.840 &7.63&  NLRG & 14.80\\  
3C\,184 &  07 39 24.31 & +70 23 10.74  & 0.994 &8.00&  NLRG & 3.45\\
3C\,200 &  08 27 25.44 & +29 18 46.51  & 0.458 &5.82&  NLRG & 3.7 \\
3C\,207 &  08 40 47.58 & +13 12 23.37  & 0.684 &7.08&  QSO& 4.1 \\
3C\,220.1& 09 32 39.65 & +79 06 31.53  & 0.61  &6.73&  NLRG & 1.8 \\
3C\,228 &  09 50 10.70 & +14 20 00.07  & 0.552 &6.42&  NLRG & 3.18\\
3C\,254 &  11 14 38.71 & +40 37 20.29  & 0.734 &7.28&  QSO& 1.90 \\
3C\,263 &  11 39 57.03 & +65 47 49.47  & 0.656 &6.96&  QSO& 1.18 \\
3C\,265 &  11 45 28.99 & +31 33 49.43  & 0.811 &7.54&  NLRG & 1.90 \\
3C\,275.1& 12 43 57.67 & +16 22 53.22  & 0.557 &6.40&  QSO& 1.99 \\
3C\,280 &  12 56 57.85 & +47 20 20.30  & 0.996 &8.00&  NLRG & 1.13 \\
3C\,292 &  13 50 41.95 & +64 29 35.40  & 0.713 &6.90&  NLRG & 2.17\\
3C\,309.1& 14 59 07.60 & +71 40 19.89  & 0.904 &7.80&  GPS-QSO& 2.30\\
3C\,330 &  16 09 34.71 & +65 56 37.40  & 0.549 &6.41&  NLRG & 2.81 \\
3C\,334 &  16 20 21.85 & +17 36 23.12  & 0.555 &6.38&  QSO & 4.24 \\
3C\,345 &  16 42 58.80 & +39 48 36.85  & 0.594 &6.66&  core-dom
QSO&1.13 \\
3C\,380 &  18 29 31.78 & +48 44 46.45  & 0.691 &7.11&  core-dom QSO& 5.67 \\
3C\,427.1& 21 04 06.38 & +76 33 11.59  & 0.572 &6.49&  LERG&  10.90\\
3C\,454.3& 22 53 57.76 & +16 08 53.72  & 0.859 &7.68&  core-dom QSO &
6.50\\
\hline
\end{tabular}
\end{center}
\end{table*}
Our analysis of the lobe emission was performed with a larger sample,
including sources at lower and higher redshift to those detailed in
Table~\ref{tab:sources}. For a full list of these sources see \citet{croston05}.

\section{The properties of the cores}
 High-frequency nuclear radio emission probes sub-arcsecond scales
of radio-loud sources, and is explained as synchrotron radiation  from the unresolved
bases of  relativistic jets, which
is anisotropic due to relativistic beaming. The  correlation found using {\it ROSAT} data between
 the nuclear, soft X-ray emission and the core radio emission of powerful radio-loud AGN 
(e.g. \citealt{hw99, brinkmann97}) suggests that at least part of the 
 soft X-ray emission is also relativistically beamed and originates at
the base of the jet. The correlation is very tight for
core-dominated quasars (CDQs), i.e. those source having jets 
pointing to small angles to the line of sight. Another
sub-class of object, lobe-dominated quasars (LDQs), were found to lie above the flux-flux
correlation valid for CDQs  \citep{dmw94}, supporting the idea that lobe-dominated quasars are viewed at an angle to the line of sight such that the observer sees in X-ray  both  the
jet-dominated component and a more isotropically emitted, probably
nucleus-related component, with the two components being more  similar
in flux density than for the CDQs. For those sources
viewed on the plane of the sky (thus strongly obscured by a torus) the
picture is less clear. \citet{dmw94} found that the core soft X-ray emission of the
galaxy 3C\,280 lies on an extrapolation of 
the correlation obtained for CDQs, and interpreted
the result as due to X-ray emission from a jet-related
component, with the nucleus-related component being obscured by a torus. 
In this picture, nucleus-related X-ray emission in radio galaxies
would be seen only at hard  ($>2.5$ keV) energies and its
characteristics have been essentially unknown before \cha\ and \xmm\ observations.

\subsection{Results}
\cha\ and \xmm\ observations allow us to answer some of the still open questions about the nature of X-ray emission from the nuclear region of high-redshift radio sources:

\begin{enumerate}
\item {\bf Do RGs have more absorbed cores than QSOs?}
Spectral analysis of 10 radio galaxies in the sample show that 70\% of
them display an absorbed nuclear component with intrinsic absorption
ranging from 0.2 to 50 $\times 10^{22}$ cm$^{-2}$. We do not find
absorption above Galactic value for any of the quasars in the sample.

\begin{figure}
\centering
\includegraphics[scale=0.46,angle=0]{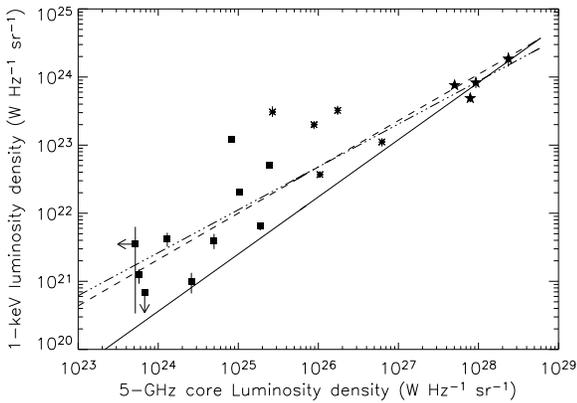}
\caption{Luminosity-Luminosity relation. Filled boxes are RGs,
asterisks are LDQs and star CDQs. Errors bars show 1$\sigma$
uncertainties. When these are not visible it is because errors are smaller than
symbols. Upper limits are at the 3$\sigma$ level. The continuous line is the
best-fit regression to the sample of CDQs only. The
dot-dashed line is the regression to the whole sample, while the
dashed line fits the sample composed of CDQs and RGs
\label{fig:lxlr}}
\end{figure}

\item {\bf If soft X-ray emission is jet-related, a correlation should
exist between core radio emission and a soft X-ray unabsorbed component of  RGs and CDQs.} 

We correlate core radio luminosity density at 5 GHz to soft, 1-keV X-ray
luminosity density of an unabsorbed component in all the sources in our sample (Figure~\ref{fig:lxlr}). We
confirm previous results found with {\em ROSAT}  about the tight
correlation valid for CDQs, and the position above the correlation for
LDQs. The 1-keV emission from the 10 RGs, which is interpreted as jet-related in CDQs, lies above the correlation valid for CDQs, suggesting that the mechanism responsible for the unabsorbed X-ray emission in the two subclasses of sources may be different.

\item   {\bf Are the two populations (RGs and QSOs) different in their X-ray spectral properties?}

Radio-loud (core-dominated or
blazar-type) quasars are found to have flatter spectral index , i.e.   
$\Gamma\sim1.5$ (e.g., \citealt{ww90},
  \citealt{brinkmann00}) in comparison to  the values of $\Gamma\sim2$ more commonly found
in their radio-quiet counterparts (e.g. \citealt{brinkmann00},
\citealt{rt00}, \citealt{galbiati05}). The flat 
 spectrum has been  interpreted as the result of beamed emission from
the jet. However, for RGs it is in principle possible to separate X-ray
emission which is  unobscured and jet-related from obscured (possibly)
accretion-related emission. From our analysis we find that the average
spectral index (Figure \ref{fig:spindex}; the absorbed component is
here considered for RGs) for all sources is $<\Gamma>$ =
$1.55\pm0.05$, for RGs alone is $<\Gamma_{RG}>$ = $1.57\pm0.03$, for CDQs $<\Gamma_{CDQ}>$ = $1.45\pm0.06$, and for LDQs $<\Gamma_{LDQ}>$ = $1.59\pm0.09$. This shows that RGs absorbed emission behaves more similarly to the spectral behaviour of radio-loud quasars (RLQs) than radio-quiet quasars (RQQs).

Interestingly, the spectral index of the RG unabsorbed X-ray emission at low energy is rather steep, with a average of $<\Gamma>$ = $2.05\pm0.25$. This is  indication that the emission mechanism responsible for the unabsorbed emission in RGs and QSOs may be different.
\begin{figure}
\centering
\includegraphics[scale=0.5,angle=0]{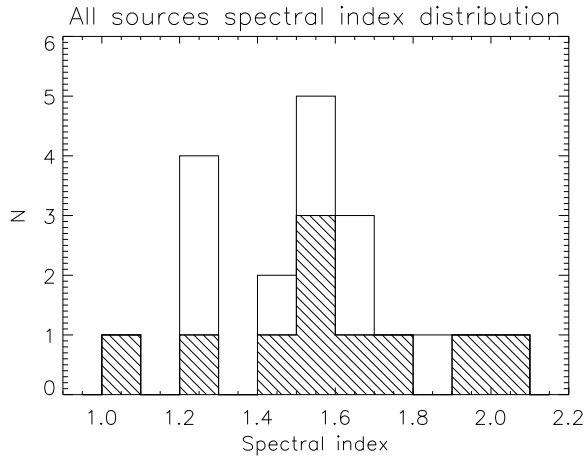}
\caption{Spectral index distribution for the whole sample. The dashed
histogram illustrates the distribution of the absorbed components of
radio galaxies, and empty boxes represent quasars.
\label{fig:spindex}}
\end{figure}
\end{enumerate}

\subsection{Conclusions}
Several results support a simple unification model to explain the X-ray emission from high-redshift radio loud sources:
\begin{itemize}
 \item RGs display higher intrinsic absorption than QSOs (as expected
if a torus is present);
\item  the unabsorbed X-ray component observed in RGs (and QSOs)
correlates with the radio core, implying that this emission is most
likely  jet related. However the steeper spectrum observed in RGs is
consistent with  synchrotron emission as the mechanism responsible for
this emission in RGs. On the other hand,

\item the flattening of the flux-flux relation (as well as the
luminosity-luminosity relation) for CDQs, i.e. the source is
under-luminous in X-ray for a given radio flux, is consistent with IC becoming dominant with decreasing angle to the line of sight;

\item  LDQs have more X-ray emission than would be expected from a jet
component alone suggesting a possible contribution to the spectrum from both jet-related and accretion-related emission mechanisms. 

\end{itemize}
The new and surprising result we find from our analysis is the flat
slope describing the absorbed X-ray spectrum of RGs. In particular its
value is flatter than that observed for RQQs. This seems to indicate
that jet emission dominates over a possible more heavily absorbed core, as verified by simulations \citep{bels06}. However, we cannot rule out that RLQs and RQQs engines are different.

\section{Extended emission from the radio lobes}
Radio synchrotron emission from radio lobes is a function of electron
density and magnetic field strength. The combination of radio
observations with resolved X-ray observations allows us to decouple these two quantities since electron density is directly measurable from the inverse Compton  (IC) emission observed in the X-rays. This may help in solving the issue of particle content  and magnetic field strength in radio lobes.

The particle content in radio galaxies and quasars is still under debate. Possible interpretations are electron-proton jets (e.g., \citealt{celfab93}) and  electron-positron jets (e.g., \citealt{wardle98,kt04}). Although relativistic protons are not directly observable by IC emission, results consistent with equipartition between the magnetic field and electron energy densities represent an indirect  means to disfavour models in which a substantial contribution to the energy density is provided by a population of protons (e.g., \citealt{mjh04, croston04}).

In this paper we summarise the results obtained from the analysis of \cha\ and \xmm\ observation of 33 classical double radio galaxies and their radio lobes. Detail of this work are described in \citet{croston05}.

\subsection{Results}
The X-ray observations available allow us to resolve emission from 38  lobes and upper limits for another 16 lobes. Of the 38 lobes with detection, 8 have sufficient counts to perform spectral analysis.

The properties of the lobes in our sample were investigated by
computing the ratio ($R$) of the observed to predicted X-ray flux at
equipartition. The predicted flux was obtained by modelling of the IC and synchrotron emission using the radio flux densities at different frequencies to normalise the synchrotron spectrum. A broken power law electron distribution with initial electron energy index $\delta=2$, $\gamma_{\rm min}$ =10 and $\gamma_{\rm max} = 10^5$, and a break energy in the range $\gamma_{\rm break} = 1200-10,000$ was used. The prediction for the CMB IC and SSC at 1 keV was determined on the basis of the modelled synchrotron spectrum for each source assuming equipartition between radiating particles and magnetic field.
In this definition $R=1$ means that the CMB and SSC model with an equipartition magnetic field and a filling factor of unity can explain the observed X-ray flux. $R>1$ indicates that either the magnetic field is lower than the equipartition value, i. e., the lobes are electron dominated, or an additional photon field is present. $R<1$ implies magnetic field domination.

\begin{figure*}
\centering
\includegraphics[scale=0.40,angle=0]{f3a.eps}
\includegraphics[scale=0.40,angle=0]{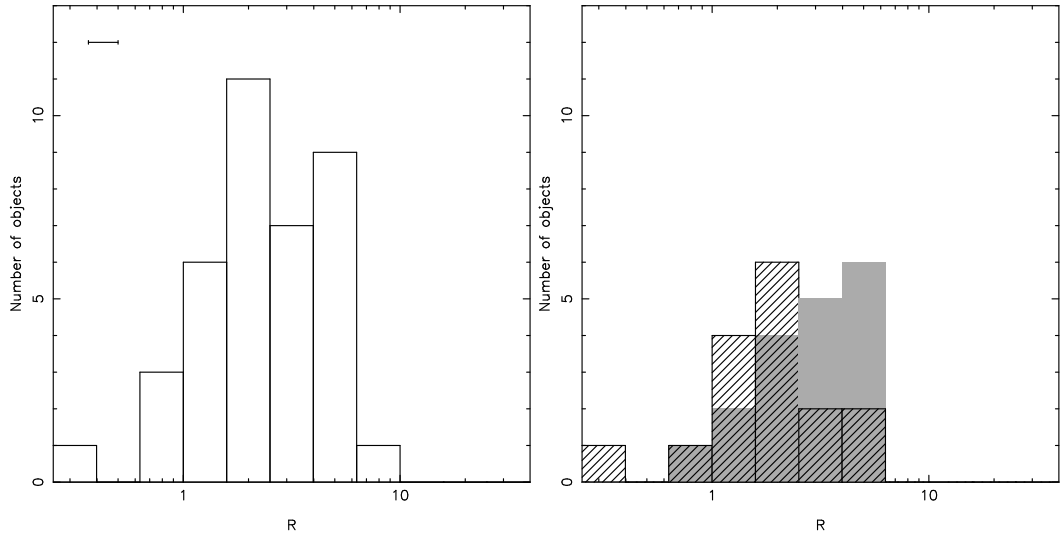}
\caption{Distribution of $R$-parameter for the detected lobe sample
(Left), and $R$ distribution of the narrow-line (hatched rectangles)
and broad line (filled rectangles) objects
discussed in \citet{croston05} (Right).
\label{fig:histolobes}}
\end{figure*}

Figure~\ref{fig:histolobes} shows the histogram of $R$ for the detected lobes. The majority of sources have $R>1$ and appear to be distributed around $R\sim2$. 

We also examined whether the observed $R$-value is related to the
type of radio source by comparing the distribution of $R$ for
narrow-line and broad-line objects. We observe that broad-line quasars
have a tendency to display higher value of $R$
(Figure~\ref{fig:histolobes}). We demonstrated \citep{croston05} that
projection effects are likely to be important in explaining the
distribution of the observed $R$-values, although  other possibilities
cannot be ruled out.

\subsection{Conclusions}
Our study shows that more than 70\% of the sources in the sample are at equipartition or electron dominated. The distribution of the $R$-values, with a peak around $R\sim2$ indicates that most of the source magnetic fields are within 35\% of equipartition, or electron dominance ($U_e/U_b$) by a factor of $\sim5$. Some sources can be magnetically dominated by at least a factor of 2. 
These results disfavour models in which FRII lobes have an
energetically dominant population of relativistic protons which are also in equipartition with the magnetic field.
 
\section{Environment of radio sources and the search of high-redshift clusters}
Because of the propagation  of radio jets, radio galaxies  are expected to lie in an external medium dense enough to confine their relativistic jets.
The combination of this hypothesis with the visibility of radio galaxies at high redshift suggested that these objects can be used as tracers of high-redshift galaxy clusters, by looking for hot intra-cluster medium around them in X-rays.

Observations with the {\em ROSAT} satellite were promising in this
context (e.g., \citealt{cf93,dmw94,crawford99,hw99}), with detection of cluster-like emission around some of the $z>0.5$ sources in the 3CRR catalogue.

\cha\ and \xmm\ make it  now possible to detect and study the environment of high redshift radio galaxies in great detail, and to separate spatially and spectrally the different components (core, lobes, jets) responsible for the X-ray emission from these sources. The picture from \cha\ and \xmm\ is rather different from that of previous satellites.

\subsection{The current picture: results from \cha\ and \xmm} 
\begin{figure}
\centering
\includegraphics[scale=0.50,angle=0]{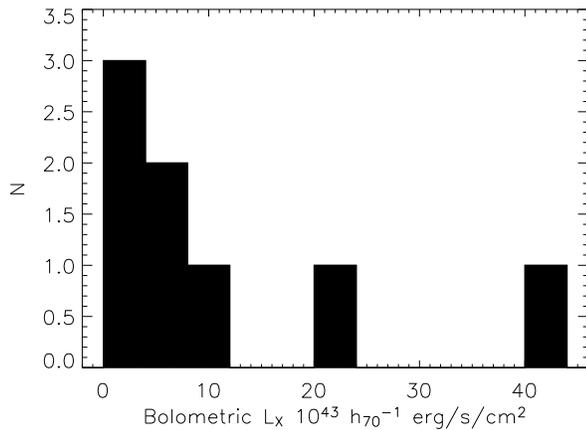}
\caption{Distribution of the bolometric X-ray luminosity of 3CRR
sources with detected extended, cluster-like emission.
\label{fig:lxhisto}}
\end{figure}

Preliminary results for sources
analysed so far are shown in Figure \ref{fig:lxhisto}. Published work
in the literature, and here assembled, shows that most of the radio galaxies
and quasars in the redshift range $0.5<z<1.0$ show a tendency to lie
in extended environments with luminosities of few $10^{43} h_{70}^{-2}$
erg s$^{-1}$ (e.g. \citealt{mjh02,cf03,ddh03}; see
Fig. \ref{fig:lxhisto}). Only few objects have been confirmed
spectroscopically (3C220.1 - \citealt{dmw01}, 3C184 and 3C292 -
\citealt{bels04}). Indeed most of the X-ray emission associated with
these and other sources was found to be elongated in the direction of the radio lobes \citep{carilli02,mjh02,ddh03,croston05}

Despite the low surface brightness emission from these objects, most of these environments are sufficient to confine the radio lobes (e.g. \citealt{mjh02,ddh03,bels04}).

The preliminary picture we observe from our study is suggesting that
radio sources at high redshift trace a particular type of
environment, i.e. they prefer poor environments at high
redshift. This also supports that they may represent a means to built
up a more unbiased sample for structures 
in the high-redshift Universe, where massive clusters are expected to be fewer in the hierarchical formation paradigm.

This calls for a systematic study of a well selected sample of radio sources, and the 3CRR catalogue is a good starting point.

Work on the detection of extended emission from the sources in Table
\ref{tab:sources} is under way. Our full study
will allow us not only to probe
the characteristics of the extended emission, but also to constrain
the physical state of the radio source itself by comparing the internal and external pressure of the radio lobes.

\section*{Acknowledgements}

We are very grateful to the organisers of this conference for allowing
us to present the work described in the paper. This well-organised meeting was enjoyable for the richness of results
presented. This paper  is based on observations obtained with \xmm, an ESA science mission with instruments and contributions directly funded by ESA Member States and the USA (NASA). This research has made use of the  NASA's Astrophysics Data System.


\begin{thebibliography}{}
\bibitem[Barthel(1989)]{barthel89} Barthel P.D., 1989, ApJ, 336, 606
\bibitem[Belsole et al.(2004)]{bels04}Belsole, E., Worrall, D.M.,
Hardcastle, M.J., Birkinshaw, M., Lawrence, C.R., 2004, MNRAS, 352,924
\bibitem[Belsole, Worrall \& Hardcastle(2006)Belsole et
al.]{bels06}Belsole,E., Worrall, D.M., Hardcastle, M.J., 2006, MNRAS, submitted
\bibitem[Brinkmann, Yuan \& Siebert(1997)Brinkmann et al.]{brinkmann97}Brinkmann W., Yuan W., Siebert J., 1997, A\&A, 319, 413
\bibitem[Brinkmann et al.(2000)]{brinkmann00}Brinkmann W.,
Laurent-Muehleisen S. A., Voges W., Siebert J., Becker R. H., Brotherton M. S., White R. L., Gregg M. D., 2000, A\&A, 356, 445
\bibitem[Brunetti et al.(2001)]{brunetti01}Brunetti, G., Cappi, M.,
Setti, G., Feretti, L., Harris, D. E., 2001, A\&A, 372, 755
\bibitem[Carilli et al.(2002)]{carilli02}Carilli, C. L., Harris, D. E., Pentericci, L., R\"ottiger, H. J. A., Miley, G. K., Kurk, J. D., van Breugel, W., 2002, ApJ, 567, 781
\bibitem[Celotti \& Fabian(1993)]{celfab93}Celotti, A. \& Fabian,
A.C, 1993 MNRAS, 264, 228
\bibitem[Crawford \& Fabian(1993)]{cf93}Crawford, C. S. \& Fabian, A. C., 1993, MNRAS, 260, 15
\bibitem[Crawford et al.(1999)]{crawford99}Crawford, C. S., Lehmann, I., Fabian, A. C., Bremer, M. N., Hasinger, G.,1999, MNRAS, 308,1159
\bibitem[Crawford \& Fabian(2003)]{cf03}Crawford, C. S. \& Fabian, A. C., 2003, MNRAS, 339, 1163
\bibitem[Croston et al.(2004)]{croston04}Croston J.H., Birkinshaw M.,
Hardcastle, Worrall D.M., 2004, MNRAS, 353, 879
\bibitem[Croston et al.(2005)]{croston05}Croston J.H., Hardcastle
  M. J., Harris D. E., Belsole E., Birkinshaw M., Worrall, D.M., 2005,
ApJ, 626,733
\bibitem[Dickey \& Lockman(1990)]{dlnh}Dickey J.M. \& Lockman F.J., 1990, ARA\&A, 28, 215
\bibitem[Donahue, Daly \& Horner(2003)]{ddh03}Donahue, M., Daly, R. A., Horner, D. J., 2003, ApJ, 584, 643
\bibitem[Galbiati et al.(2005)]{galbiati05}Galbiati E., et al., 2005,
  A\&A, 430, 927
\bibitem[Hardcastle \& Worrall(1999)]{hw99}Hardcastle, M. J., Worrall, D. M., 1999, MNRAS, 309, 969
\bibitem[Hardcastle et al.(2001)Hardcastle, Birkinshaw \&
Worrall]{hbw01} Hardcastle M. J., Birkinshaw M., Worrall D. M., 2001, MNRAS,
326, 1499
\bibitem[Hardcastle et al.(2002)]{mjh02}Hardcastle, M. J., Birkinshaw, M., Cameron, R. A., Harris, D. E., Looney, L. W.,  Worrall, D. M., 2002, ApJ, 581, 948
\bibitem[Hardcastle et al.(2004)]{mjh04}Hardcastle, M.J., Harris,
D.E., Worrall, D.M., Birkinshaw, M., 2004, ApJ, 612, 729
\bibitem[Kino \& Takahara(2004)]{kt04}Kino, M., \& Takahara, F., 2004,
MNRAS, 349, 336
\bibitem[Laing et al.(1983)Laing, Riley \& Longair]{3crrcat}Laing R. A., Riley
J. M., Longair M. S., 1983, MNRAS, 204, 151
\bibitem[Marshall et al.(2005)]{marshall05}Marshall H. L., et al., 2005, ApJS, 156, 13
\bibitem[Overzier et al.(2005)]{overzier05}Overzier, R. A., Harris,
D. E., Carilli, C. L., Pentericci, L., R\"ottgering, H. J. A., Miley,
G. K., 2005, A\&A, 433, 87
\bibitem[Reeves \& Turner(2000)]{rt00}Reeves J.N. \& Turner M.J.L.,
2000, MNRAS, 316, 234
\bibitem[Urry \& Padovani(1995)]{UrryPadovani95}Urry C. M. \&
Padovani P., 1995, PASP, 107, 803
\bibitem[Wardle et al.(1998)]{wardle98}Wardle, J.F.C, Homan, D.C.,
Ojha, R., Roberts, D.H., 1998, Nature, 395, 457
\bibitem[Worrall \& Wilkes(1990)]{ww90}Worrall D.M. \& Wilkes B.J., 1990, ApJ, 360, 396
\bibitem[Worrall et al.(1994)]{dmw94}Worrall, D. M., Lawrence, C. R., Pearson, T. J., Readhead, A. C. S., 1994, ApJ, 420, L17
\bibitem[Worrall et al.(2001)]{dmw01}Worrall, D. M., Birkinshaw, M., Hardcastle, M. J., Lawrence, C. R., 2001, 326, 1127
\end{thebibliography}
\end{document}